\documentclass[aip,jcp,preprint]{revtex4-2}

\usepackage{mathptmx}
\usepackage{graphicx} 
\usepackage{float}
\usepackage{fancyhdr}
\usepackage{hyperref}
\usepackage{natbib}
\usepackage{bibentry}
\usepackage{relsize}
\usepackage{array}
\usepackage{booktabs}
\usepackage{float}
\usepackage{subcaption} % For subfigures
\usepackage[dvipsnames]{xcolor}
\usepackage[table]{xcolor}
\usepackage[a4paper, total={6.5in, 8in}]{geometry}
\newcommand{\bse}{\begin{subequations}}
\newcommand{\ese}{\end{subequations}}

\begin{document}
\bibliographystyle{tim}
\nobibliography*

\title{A Computational Study of Organic Molecular Crystals for Photocatalytic Water Splitting}
\author{James D. Green}
\affiliation{Department of Chemistry, Molecular Sciences Research Hub, Imperial College London, London W12 0BZ, UK}
\author{Daniel G.  Medranda}
\affiliation{Department of Physics, Imperial College London, London, UK}
\author{Hong Wang}
\affiliation{Department of Chemistry, University of Liverpool, Liverpool L69 7ZD, United Kingdom}
\author{Andrew I. Cooper}
\affiliation{Department of Chemistry, University of Liverpool, Liverpool L69 7ZD, United Kingdom}
\author{Jenny Nelson}
\affiliation{Department of Physics, Imperial College London, London, UK}
\author{Kim E. Jelfs}
\affiliation{Department of Chemistry, Molecular Sciences Research Hub, Imperial College London, London W12 0BZ, UK}
\email{j.green@imperial.ac.uk}

\date{\today}

\maketitle
\section*{Abstract}

Organic crystalline materials are potential candidates for photocatalytic overall water splitting (OWS). Although organic crystals have been heavily investigated for application in organic electronics, such as organic light-emitting diodes (OLEDs) and solar cells, there have been comparatively fewer studies into OWS in these materials. A major challenge is the large number of electronic and structural criteria that must be met for a material to make a viable OWS photocatalyst. Optical absorption, reduction and oxidation potentials and charge-transport properties are among the key considerations, and these are influenced both by molecular properties and the solid-state packing arrangement, making computational modelling challenging. Here, we investigate a series of known organic electronic materials that have published crystal structures using periodic density functional theory (DFT) and compare their calculated electronic properties of optical absorption and reduction and oxidation potentials with literature experimental data. Furthermore we perform a series of gas-phase molecular calculations which show a good agreement with literature data and periodic DFT for the optoelectronic properties of the organic molecular crystals studied, showing that gas-phase molecular calculations could be used to screen organic crystals for OWS at a reduced computational cost.
\clearpage

\section*{Introduction}

Recent years have seen a progressive shift towards organic materials in optoelectronic applications, not only offering a more sustainable and cost effective alternative to conventional inorganic materials, but also offering new functionality.\cite{hon21a,ada01a,tun05a,uoy12a,cla10a,cla10b,zir15a,son21a,che18a} For example, organic light-emitting diodes (OLEDs) has taken over the market for high-end displays, and organic photovoltaics (OPV) have found a useful niche for indoor devices among other applications.\cite{hon21a,ada01a,uoy12a,hal95a,cla10b,zha22a,yoo04a,son21a} Photocatalysts are yet another industrially relevant class of optoelectronic materials, spanning a wide range of different reactions from removing air pollutants to producing chemical feedstocks and fuels using sunlight.\cite{che18a, che10a, wan19a,abd21a} As well as for OLEDs and OPV, photocatalytic processes have also benefitted from ever-improving organic catalysts.\cite{che18a} One highly promising photocatalytic process for the future energy economy is solar-driven overall water splitting (OWS) reaction,  where water is converted into hydrogen and oxygen by the aid of sunlight and a photocatalyst, without requiring any sacrificial agents.\cite{che17a,zha17a,wan19a,che24a} OWS consists of two half-reactions: the hydrogen evolution reaction (HER), where 2H$^+$ is reduced to H$_2$, and the oxygen evolution reaction (OER), where H$_2$O is oxidised to $\frac{1}{2}$O$_2$ and 2H$^+$.

The best currently known photocatalysts for OWS are crystalline inorganic materials, for example NaTaO$_3$,\cite{kat03a} Ga$_2$O$_3$,\cite{wan12a} and SrTiO$_3$\cite{kat02a} have all been widely studied as photocatalysts for OWS. There have also been sustained efforts to find suitable and competitive organic photocatalysts for OWS, however, thus far the solar hydrogen and oxygen production rates of organic materials have fallen short of those demonstrated by inorganic materials.\cite{spr15a,liu15a,ait20a,li22a,che24a} Despite this, some promising materials have been identified, with efforts focused primarily on extended organic materials, such as such as carbon nitrides,\cite{mae09a,liu15a,zha17a} and 2D/3D covalent organic frameworks\cite{li22a,che24a}. More recently, HER has also been demonstrated in organic molecular crystals but OWS has yet to be achieved in these systems.\cite{ait20a,guo22a,yu24a} One of the fundamental challenges is that very few organic materials have the correct energetic properties to simultaneously catalyse HER and OER, although Z-scheme systems in which two complementary materials are combined to simultaneously catalyse OER and HER offer a solution to this challenge.\cite{qin21a,she23a} Given the chemical and structural diversity of organic molecular crystals, and also that comprehensive studies of such materials for OWS are relatively sparse in the literature, this prompted a focused computational investigation into the electronic properties of such materials for OWS.

To propose suitable materials for OWS with the necessary electronic properties, we must examine the initial steps of the OWS reaction, namely photoexcitation of the catalyst, charge separation, charge-carrier transport (to the interface of the catalyst and water) and charge injection (of electrons and holes into protons and water respectively). Firstly, it is fairly obvious that many materials can be ruled out as photocatalysts based on their inability to absorb light in the solar spectrum. Secondly, even after successful photoexcitation, it is by no means a given that sufficient charge separation also occurs. Thirdly, provided that charge-separation does occur, sufficient numbers of charge carriers will need to make it to the surface of the photocatalyst material before recombining. Fourth and finally, the electrochemical potential of the photogenerated electrons and holes still needs to be sufficient to allow for the injection of charges into the protons and water. Therefore the electronic properties and photophysics of the photocatalyst material itself is of primary importance to allow adequate HER and OER performance.\cite{che17a,zha17a,wan19a,che24a} 

The fourth step, injection of charges, is more complicated, as in reality it is necessary to use a co-catalyst that acts as an intermediary between the catalyst and protons/water to reduce the activation energy of these charge transfer reactions and allow OWS to occur at a reasonable rate.\cite{zha17a} In practice, separate HER and OER cocatalysts are used to accept electrons and holes respectively from the photocatalyst and inject them into the protons and water, however, consideration of cocatalsyts does not ultimately change the energetic criteria that electrons and holes must provide sufficient driving force for the OER and HER reactions.\cite{zha17a} Here, we focus on the optical absorption and reduction/oxidation potentials of molecular electronic materials. We will also briefly introduce charge-transport in molecular crystals but will leave discussion of charge separation aside.

Fortunately, theoretical models and computational calculations allow us to interrogate the fundamental electronic properties of materials. Quantum chemical calculations using density functional theory (DFT) are a particularly popular approach and can provide accurate insight into such properties.\cite{g16,orca,vasp,van15a,nee06a,jai16a,pas10a} These calculations are grouped into three main types based on the description of the chemical system; i) solid-state DFT calculations describing the system as a bulk material using periodic boundary conditions (PBC)\cite{vasp,hut14a}; ii) molecular DFT calculations describing the system as an isolated molecule or molecules,\cite{g16,orca,hut14a} either in the gas-phase or in solution; and calculations of materials described by large clusters of molecules not imposing periodic boundary conditions but instead employing interacting quantum mechanics (DFT) and molecular mechanics regions (QM/MM)\cite{sve96a,she03a}. We will refrain from discussing the third type of calculations any further, but one should be aware of the power of QM/MM approaches for treating large systems lacking periodicity or with very large unit cells.\cite{sve96a,she03a} With relevance to OWS, three key types of descriptors can be calculated for materials that relate to their performance and suitability as photocatalysts:

\paragraph{Optical absorption:}
The excited states of a material can be calculated using linear response methods such as time-dependent DFT (TD-DFT). The energy of the first excited state ($S_1$) is often used as a reference point, as this is the lowest energy photon that can be absorbed via an electronic transition in the material.\cite{lop17a,shi25a,hel19a} Additionally, the full UV-visible spectrum of the material may be simulated taking into account all calculated states and their oscillator strengths, providing a clearer picture of which electronic states will be responsible for absorption across the UV-visible spectrum, especially important when $S_1$ has a vanishingly small oscillator strength.\cite{hel19a,gre22a} 
 
\paragraph{Reduction and oxidation potentials:}
Ionisation potential (IP) and electron affinity (EA) are key to determining whether the photogenerated holes and electrons have sufficient electrochemical potential to drive OER and HER respectively. The IP and AE are related to the valence band maximum (VBM) and conduction band minimum (CBM) positions respectively in the bulk system, or highest occupied molecular orbital (HOMO) and lowest unoccupied molecular orbital (LUMO) energies in an isolated molecule. The IP must be greater than the electrochemical potential of the water oxidation reaction (1.23 V, 5.65 V relative to the vacuum level) and the EA must be lower than the  electrochemical potential of the proton reduction reaction (0 V, 4.42 V relative to the vacuum level).\cite{faw08a,zha18a,wan20a,she24a} 

\begin{figure}[h]
    \centering
    \includegraphics[width=0.8\textwidth]{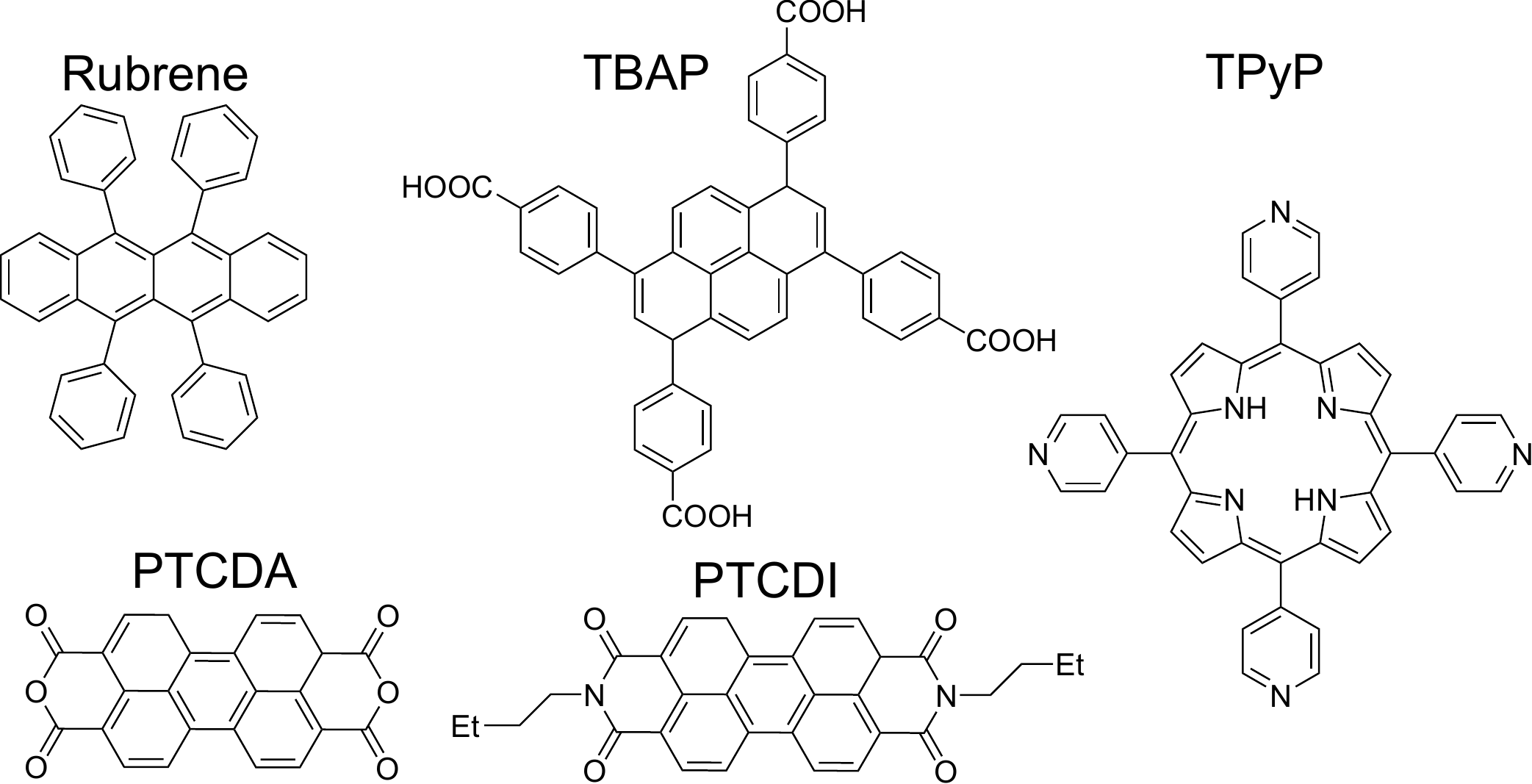}
    \caption{The five molecular crystals studied here:  rubrene, 4,4',4'',4'''-(Pyrene-1,3,6,8-tetrayl)tetrabenzoic acid (TBAP), perylenetetracarboxylic dianhydride (PTCDA), dibutyl-perylenetetracarboxylic diimide (sometimes referred to as C4-PTCDI and referred to here as simply PTCDI) and 5,10,15,20-tetra(4-pyridyl)-porphine (TPyP). }
    \label{fig:mols}
\end{figure}

\paragraph{Charge transport:}
Though we do not numerically evaluate the charge transport properties of the organic molecular crystals here, as an aside we will briefly introduce some concepts of charge transport.

The effectiveness of charge carrier (electron and hole) transport through a material is usually quantified by the carrier mobility, $\mu$, normally measured in units of cm$^2$ V$^{-1}$s$^{-1}$. $\mu$ is an empirical quantity, but can be estimated by first-principles simulations.\cite{nel09a,rod10a} These simulations are based on either a delocalised, band-like model or a localised, hopping-like model of charge transport. The choice of model depends on the nature of the system under investigation, where for very delocalised periodic materials the band-like model is most appropriate, and for very localised, molecular systems especially those with low translational symmetry, the molecular hopping-like model is most appropriate. These two models, however, are limiting cases and in many organic materials, aspects of either model apply.\cite{yav17a} 

In the case of band-like transport, charge carriers are completely delocalised across the material and move continuously in bands. Band-like transport is hindered by structural defects and thermal vibrations (phonons) resulting in a negative temperature coefficient i.e. decrease in mobility with increasing temperature. %, although the effect of such defects and vibrations (phonons) will not be discussed further here. 
$\mu$ can be shown to depend on the effective carrier masses $m^*$ at the band edges and scattering times $\tau$ due to thermal noise and defects according to the band-like model $\mu=\frac{e\tau}{m^*}$.\cite{sch14a,mar20a,hu15a,luo24a,phu18a} 

In the opposite extreme, the wavefunctions of charge carriers are completely localised to individual sites in the material and transport occurs via discrete, thermally activated hops between sites, which in this case are individual molecules. This hopping-like transport is characterised by an initial increase in mobility with increasing temperature up until an optimum after which mobility decreases again. The hopping-rate $\Gamma_{ij}$ between individual pairs of molecules $i$ and $j$, a fundamental property, can be expressed in a Marcus model $\Gamma_{ij}=\frac{J_{ij}^2}{\hbar}\sqrt{\frac{\pi}{\lambda k_BT}}\exp(\frac{-\lambda}{4k_BT})$, where $\lambda$ is the reorganisation energy --- usually considered to be an intrinsic molecular property, $J_{ij}$ the transfer integral ---  a property linked to pairs of molecules, $k_B$ the Boltzmann constant, $\hbar$ the reduced Planck constant and $T$ the temperature. This model assumes all molecules are identical, there is no applied electric field and that the carrier concentration is low. The mobility can then be simulated by calculating $\Gamma_{ij}$ for all pairs of molecules in the material and either directly solving the master equation, applying Monte Carlo simulations or approximated from the diffusion coefficients $D_\alpha $ in each cell direction $\alpha$ using the Einstein relation $\mu=\frac{qD}{k_BT}$.\cite{sch14a,nel09a,rod10a,che16a, ma17a,yin08a,yav17a} 

The quantities $m^*$ and $\tau$ (band-like model) and $J$, $\lambda$ and $\Gamma$ (hopping-like model) can be theoretically predicted by DFT calculations.\cite{yav17a,nel09a} Both $m^*$ and $J$ are shown to weakly correlate with the experimentally measured mobility --- a decrease in $m^*$ or increase in $J$ is usually expected to result in a higher $\mu$. Furthermore, for molecular systems, the value of $2J/\lambda$ --- a qualitative approximation for the electron-phonon coupling strength, can be used to roughly categorise transport as band-like when $2J/\lambda\gg1$, or hopping-like when $2J/\lambda\ll1$.\cite{yav17a}

\paragraph{Aim of this article}
Here, we consider five organic molecular crystals presented in Figure~\ref{fig:mols} that have all been previously investigated for (opto)electronic devices, namely: rubrene, 4,4',4'',4'''-(Pyrene-1,3,6,8-tetrayl)tetrabenzoic acid (TBAP), perylenetetracarboxylic dianhydride (PTCDA), perylenetetracarboxylic diimide (PTCDI) and 5,10,15,20-tetra(4-pyridyl)-porphine (TPyP).\cite{tet06a,par08a,bal20a,tsa20a,shi09a,zha22b,ait20a,yu24a,guo22a} The pyrene- and perylene-based molecules (TBAP, PTCDA, PTCDI) and their derivative compounds have also been of interest for photocatalytic HER and OER.\cite{ait20a,yu24a,guo22a} We apply two of the aforementioned DFT-based descriptors, optical absorption and reduction/oxidation potentials, to these five organic molecular crystals and discuss their potential for OWS. Furthermore, we compare descriptors calculated using solid-state calculations using periodic boundary conditions to less computationally demanding molecular calculations to test the possibility of screening candidate organic molecular crystals for OWS using less intensive calculations. 

\section*{Methods}
\subsection*{Computational crystal structures}
The following X-ray crystal structures were taken from the Cambridge Structural Database with reference codes and space groups in parentheses: rubrene (QQQCIG01, $Cmca$), TBAP (MUDSIQ, $C2/m$), PTCDA (SUWMIG02, $P21/n$), C4-PTCDI (DICNUY02, $P\text{-}1$) and TPyP (COWSIR, $C2/c$).\cite{gro16a} For each of the structures, the primitive cell and atomic positions were relaxed to their theoretically optimal values using periodic DFT in the Vienna Ab-initio Structure Package version 6.5 (VASP).\cite{vasp} For each structure, a sufficiently large k-point mesh was selected such that the total energy with respect to increasing the number of k-points had converged to within 1 eV/atom. The number of k-points in respective $a$ x $b$ x $c$ cell directions used for rubrene, TBAP, PTCDA PTCDI and TPyP were 2x2x2, 3x3x3, 3x3x3, 3x3x3, and 2x2x2 respectively. The theoretical model employed the Perdew-Burke-Ernzerhof (PBE) density functional with the damped dispersion correction of Becke, Johnson and Grimme (for non-covalent interactions) and used a plane-augmented-wave (PAW) basis with an energy cutoff (ENCUT) of 650 eV.\cite{per96a,gri11a} The calculations were deemed to have converged to a relaxed structure when the force acting on all atoms fell below 0.01 eV \AA$^{-1}$. The root mean squared deviation of atomic positions for a cluster of 30 molecules (RMSD$_{30}$) between the experimental and DFT relaxed crystal structures were below 0.3 {\AA} for all molecular crystals. The relaxed crystal structures are shown below in Figure~\ref{fig:crystals}.

\begin{figure}[h]
    \centering
    \includegraphics[width=0.75\textwidth]{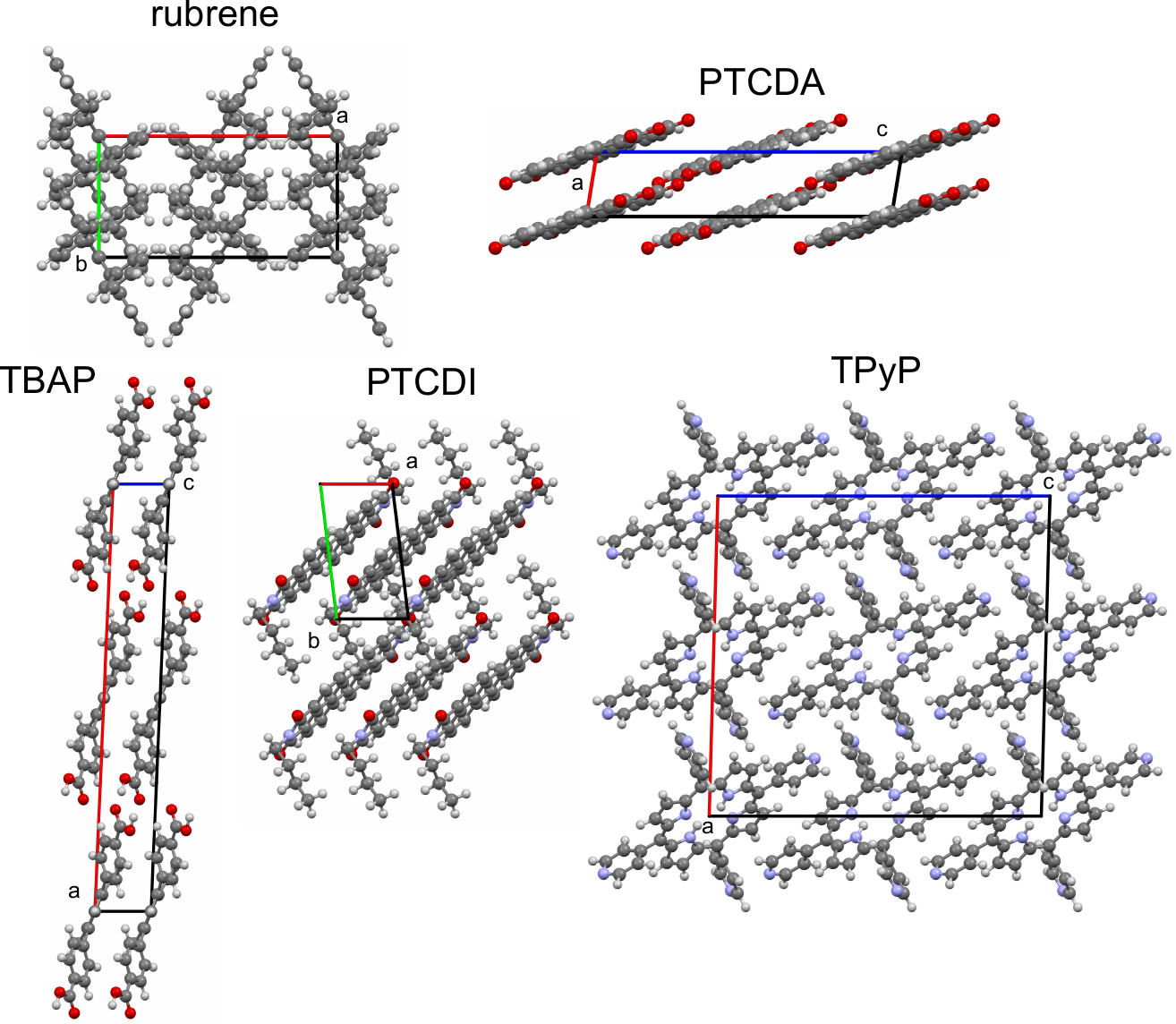}
    \caption{DFT relaxed crystal structures of five organic crystals: rubrene (viewed perpendicular to $c$ axis), b) TBAP (viewed perpendicular to $b$ axis), c) PTCDA (viewed perpendicular to $b$ axis), d) PTCDI (viewed perpendicular to $c$ axis), e) TPyP (viewed perpendicular to $b$ axis). C atoms coloured in grey, H in white, N in blue and O in red. Figure produced using Mercury 2024.1.0.\cite{mac20a}} %The directions of extended $\pi$-stacking interactions are shown by dotted lines. }
    \label{fig:crystals}
\end{figure}

\subsection*{IP and EA calculations}
The computational band structures of bulk crystals were simulated using VASP\cite{vasp} on the PBE-D3 relaxed structures using the HSE06 hybrid density functional and ENCUT of 650 eV with the following k-point meshes for rubrene, TBAP, PTCDA, PTCDI and TPyP respectively: 2x2x2, 1x1x4, 3x1x1, 4x2x1, 2x2x2. The valence band maximum (VBM) and conduction band minimum (CBM) were extracted from the band structures - rubrene and PTCDI have direct band gaps of 1.63 eV and 1.38 eV, whereas TBAP, PTCDA and TPyP have indirect band gaps of 2.26 eV, 1.83 eV and 2.11 eV respectively. 

Vacuum slabs were constructed for the organic crystals with a bulk thickness of 20 \AA\hspace{0.6mm} and vacuum gap of 40 \AA\hspace{0.6mm} starting from the relaxed cells. The electrostatic potential was calculated by a self-consistent calculation on the vacuum slabs with HSE06/ENCUT = 650 eV. The surface dipole method was then used to correct the VBM and CBM for the electrostatic potential inside the bulk crystal to give ionization potentials (IP) and electron affinities (EA) for each crystal: $\mathrm{IP}=\epsilon_\mathrm{SD}-\epsilon_\mathrm{VBM}$, $\mathrm{EA}=\mathrm{IP}+\epsilon_g$ where $\epsilon_\mathrm{SD}$ is the surface dipole energy, $\epsilon_\mathrm{VBM}$ is the VBM energy and $\epsilon_g$ is the band gap. 

\subsection*{Excitation energies of bulk systems}
    The periodic time-dependent density functional perturbation theory (TD-DFPT) method implemented in the CP2K program version 2024.3 (CP2K) using the Tamm-Dancoff approximation was used to calculate the excited states of the molecular crystals.\cite{cp2k,ian05a} TD-DFPT calculations were performed on supercells of the relaxed structures with minimum lattice constants of 13 \AA. The first thirty roots of the excited state Hamiltonian were calculated which is a sufficiently large number to ensure that the correct $S_1$ state was captured for all molecules due to state energy reordering. The M06-2X\cite{zha08a} and $\omega$B97x\cite{cha08a} functionals were employed with a gaussian augmented-plane-wave basis (GAPW) using the DZVP-MOLOPT valence basis and GTH-PBE pseudopotential (for core electrons) for all atoms. 

\subsection*{Molecular excitation and orbital energies}
DFT calculations in Orca 6.0\cite{orca,orca5,nee02a,nee03a,nee09a,nee23a} were performed for molecular HOMO/LUMO energies using a range of functionals (PBE, B3LYP, $\omega$B97x, LC-PBE and M06-2X)\cite{per96a, bec93a,cha08a,iik01a,zha08a} and the def2-TZVP basis set. The excited states of single molecules were subsequently calculated using standard TD-DFT in Orca 6.0 using the same functionals and basis set.\cite{run84a,orca5} The first ten roots of the excited state Hamiltonian were calculated.

\section*{Results and Discussion}
\subsection*{Optical gap}
\begin{figure}[h]
    \centering
    \includegraphics[width=0.6\textwidth]{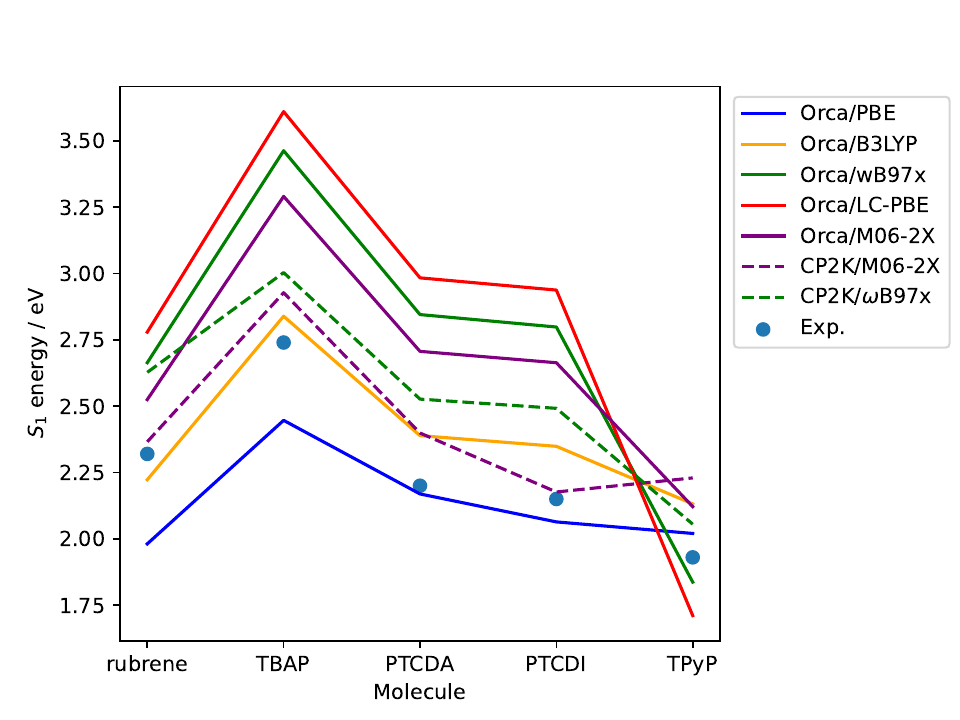}
    \caption{Comparison of the $S_1$ energies for five organic molecular crystals calculated by periodic TD-DFPT (M06-2X and $\omega$B97x in CP2K\cite{cp2k}) with molecular TD-DFT (using four functionals: B3LYP, $\omega$B97x, LC-PBE and M06-2X in Orca\cite{orca,orca5}). The experimental optical gap deduced from thin-film UV-Visible absorption spectra is also plotted for reference.\cite{irk12a,yam12a,fer06a,mak19a}}
    \label{fig:dft_s1}
\end{figure}

The first excited state ($S_1$) energies of the five compounds calculated by periodic TD-DFPT using M06-2X are shown in Table~\ref{tab:s1} and also are plotted in Figure~\ref{fig:dft_s1}. These calculations predict that all molecules have an optical gap within the visible region (400 - 700 nm). When compared to the optical gaps taken from thin-film UV-visible absorption spectra\cite{irk12a,yam12a,fer06a,all09a,mak19a}, the overall accuracy of periodic TD-DFPT is reasonably good for all molecules predicting optical gaps within 0.2 eV of the experimental values for rubrene, TBAP, PTCDA and PTCDI and within 0.3 eV for TPyP. However, as evident from Figure~\ref{fig:dft_s1} the trend in the $S_1$ energies from periodic M06-2X deviates from that of experiment, as TPyP which has the lowest experimental $S_1$ energy is predicted to have the second lowest by periodic M06-2X. TD-DFPT calculations were repeated with periodic $\omega$B97x which correctly reproduces the experimental trend in the $S_1$ energies at the cost of some accuracy. Here the $S_1$ energy of TPyP is stabilised by around 0.2 eV, but conversely the $S_1$ energies of all other molecules are destabilised by a similar amount in $\omega$B97x relative to M06-2X. These results are also shown in Table~\ref{tab:s1}. 

\begin{table}[h]
    \centering
    \caption{$S_1$ energies calculated for a series of organic molecular crystals in TD-DFPT using periodic M06-2X/DZVP and periodic $\omega$B97x/DZVP. The experimental optical gaps measured in the solid state are also included with their corresponding literature reference provided.}
    \begin{tabular}{c|c|c|c|c|c|c}
         Molecule & $S_1$(M06-2X) / eV& $S_1$($\omega$B97x) / eV & $S_1$(exp.) / eV& Ref.  & Error (M06-2X) / eV & Error ($\omega$B97x) / eV \\ \hline
         Rubrene & 2.37 &	 2.63 & 2.32 &\citenum{irk12a}&0.05&0.31\\
         TBAP & 2.93&3.00 & 2.74 & \citenum{yam12a}&0.18 & 0.26 \\
         PTCDA &2.40&	2.53 & 2.2 & \citenum{fer06a}& 0.20&0.33\\
         PTCDI & 2.18& 2.49& 2.2 &\citenum{all09a}& $-$0.02&0.29\\
         TPyP & 2.23	&	2.06 &1.93 & \citenum{mak19a}& 0.30 &0.12
    \end{tabular}
    \label{tab:s1}
\end{table}

We now compare the $S_1$ energies predicted by periodic TD-DFPT to those obtained using molecular TD-DFT calculations as shown in Figure~\ref{fig:dft_s1}. Five different exchange-correlation functionals: PBE,\cite{per96a} B3LYP,\cite{bec93a} $\omega$B97x,\cite{cha08a} LC-PBE\cite{iik01a} and M06-2X\cite{zha08a} were tested to ascertain the functional dependence of the excitation energies. These results are presented in Figure~\ref{fig:dft_s1}. The molecular calculations using range-separated and hybrid meta-GGA $\omega$B97x (RMSE = 0.53 eV), LC-PBE (RMSE = 0.66 eV) and M06-2X (RMSE = 0.41 eV) functionals all perform worse than periodic M06-2X (RMSE = 0.18 eV) and periodic $\omega$B97x (RMSE = 0.27 eV) at predicting $S_1$ and generally overestimate $S_1$ energies by a large margin (apart from for TPyP, RMSE is the root mean squared error). On the other hand, when using the hybrid GGA B3LYP functional (RMSE = 0.15 eV), molecular calculations perform on a par with the periodic calculations and closely reproduce the experimental $S_1$ energies. The reason for the range-separated and hybrid meta-GGA functionals, usually considered to be higher in accuracy, performing worse than B3LYP is due to the well known cancellation of two types of errors in the molecular calculations: B3LYP tends to systematically underestimate the HOMO-LUMO/band gap and thus the excitation energies, but conversely, excitation energies tend to be systematically lower in the solid state compared to the gas phase due to a combination of exciton delocalisation and electrostatic polarisation effects of the solid medium.\cite{sch11a,kos25a,hsi25a}

To assess the relative computational cost of periodic versus molecular calculation of the optical gap, a periodic calculation (TD-DFPT with a single root using $\omega$-B97x/DZVP-MOLOPT in CP2K on a 512 atom supercell of PTCDI) was compared with a molecular calculation (TD-DFT B3LYP/def2-TZVP with a single root in Orca on a single PTCDI molecule) on the same sixteen-core desktop computer, where the former took four hours and the latter two and a half minutes, making the molecular calculation two orders of magnitude faster. Although B3LYP should not be relied upon in the case of systems with long-range charge transfer states where it tends to be particularly unreliable,\cite{cha08a,tor16a} it is promising that a computationally inexpensive molecular TD-DFT approach may suffice for predicting the solid-state optical absorption of simple organic molecular crystals.

\subsection*{Reduction and oxidation potentials}
\begin{figure}[h]
    \centering
    \includegraphics[width=0.6\textwidth]{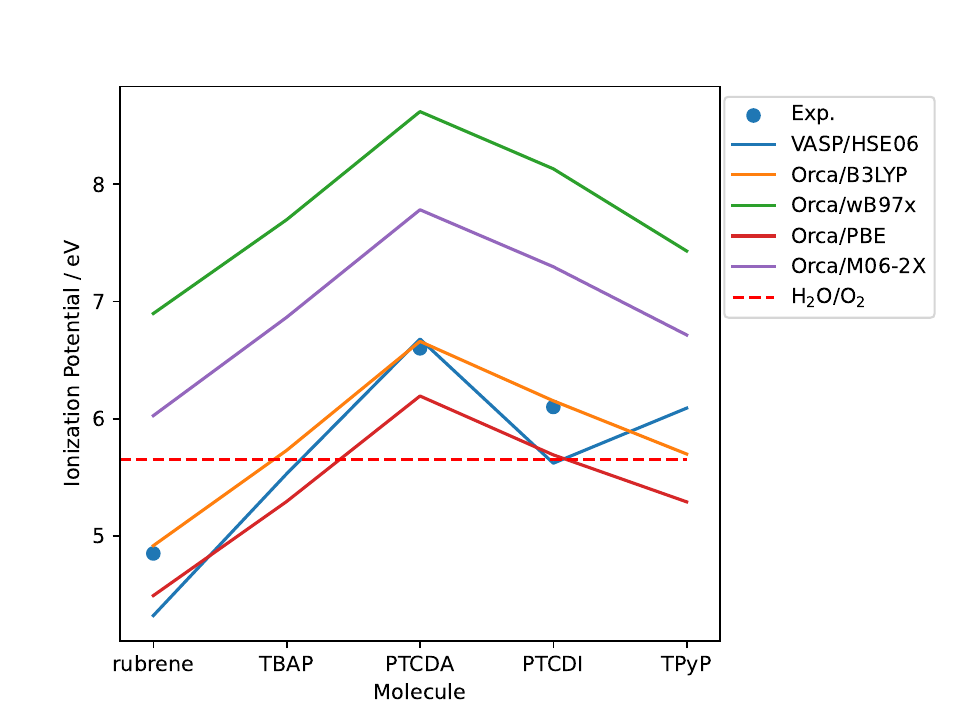}
    \caption{The ionisation potentials for five organic molecular crystals calculated by HSE06 PAW compared with values calculated by molecular DFT (using four functionals: B3LYP, $\omega$B97x, LC-PBE and M06-2X). The experimental IP is also plotted (where available). The water oxidation potential is also shown for reference.\cite{nak08a,sch11a,bal20a}}
    \label{fig:ips}
\end{figure}

\begin{figure}[h]
    \centering
    \includegraphics[width=0.6\textwidth]{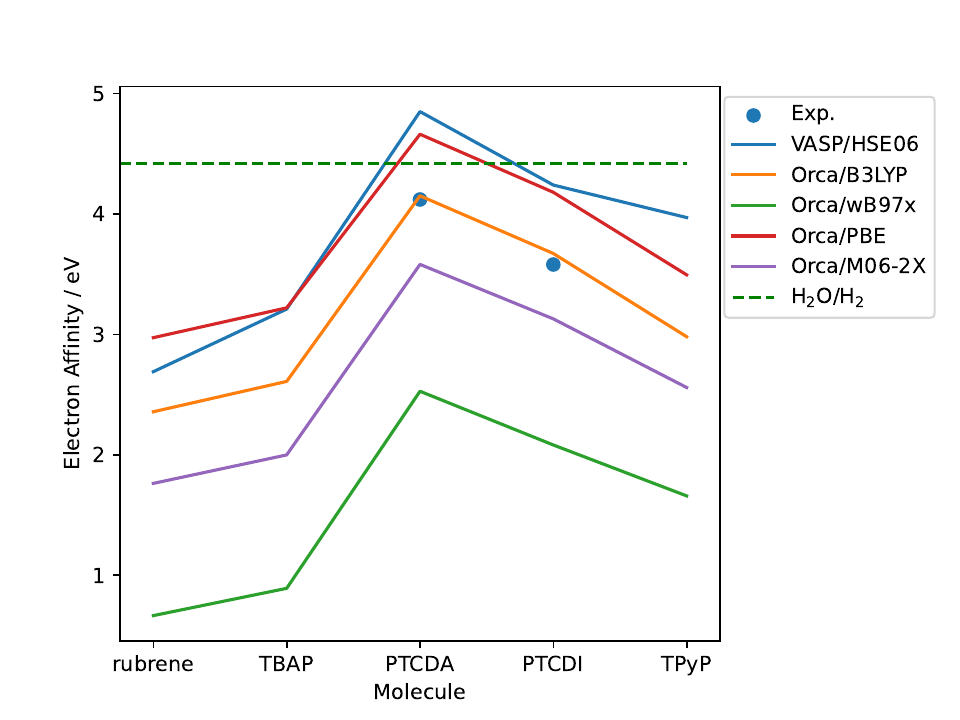}
    \caption{The electron affinites for five organic molecular crystals calculated by HSE06 PAW compared with values calculated by molecular DFT (using four functionals: B3LYP, $\omega$B97x, LC-PBE and M06-2X). The experimental EA is also plotted (where available). The proton reduction potential is also shown for reference.}
    \label{fig:eas}
\end{figure}

The ionisation potentials (IP) and electron affinities (EA) were calculated for the organic crystals using HSE06 PAW and the vacuum slab surface dipole method as described in the previous section. To complement these results, molecular DFT calculations with a range of functionals (PBE, B3LYP, $\omega$B97x, LC-PBE, M06-2X) were employed to estimate the IP (by taking the negative of the HOMO energies) and the EA (by taking the negative of the LUMO energies).\cite{per96a, bec93a,cha08a,iik01a,zha08a} These results are shown in Tables~\ref{tab:ip} and \ref{tab:ea} and Figures~\ref{fig:ips} and \ref{fig:eas}. We also draw reference to the oxidation potential of water (5.65 V) and reduction potential of protons (4.42 eV) to compare the predicted driving force for HER and OER for the molecular crystals. 
\begin{table}[h]
    \centering
     \caption{Ionisation potentials (IP) calculated using HSE PAW, and with PBE, B3LYP and $\omega$B97x on single-molecules for a series of five organic molecular crystals. Literature experimental values are also shown where available for comparison with their corresponding citation *(experimental value for PTCDI quoted here is for C12-PTCDI, the calculated value is for C4-PTCDI).}
    \begin{tabular}{c|c|c|c|c|c|c}
         Molecule & IP(exp.) / eV &Ref.& IP(HSE06 PAW) / eV & IP(PBE) / eV & IP(B3LYP) / eV &  IP($\omega$B97x) / eV \\ \hline
         Rubrene & 4.85 &\citenum{nak08a} & 4.32 &4.46& 4.81&6.83\\
         TBAP & -& - &5.53&5.19 &5.47 &7.44 \\
         PTCDA & 6.6 &\citenum{sch11a} &6.68 &6.24 &6.63 &8.65\\
         PTCDI & 6.1* & \citenum{bal20a}&5.62 &5.72&6.12& 8.14\\
         TPyP & -& -& 6.09 &5.26  & 5.57 & 7.34
    \end{tabular}
    \label{tab:ip}
\end{table}

\begin{table}[h]
    \centering
     \caption{Electron affinities (EA) calculated using HSE PAW and with PBE, B3LYP and $\omega$B97x on single-molecules for a series of five organic molecular crystals. Literature experimental values are also shown where available for comparison  with their corresponding citation *(experimental value for PTCDI quoted here is for C12-PTCDI, the calculated value is for C4-PTCDI).}
    \begin{tabular}{c|c|c|c|c|c|c}
         Molecule & EA(exp.) / eV&Ref.& EA(HSE06 PAW) / eV & EA(PBE) / eV & EA(B3LYP) / eV &  EA($\omega$B97x) / eV \\ \hline
         Rubrene & -& -& 2.69&2.98 &2.27&0.63\\
         TBAP & -& -&3.21 &3.34& 2.61& 1.00\\
         PTCDA & 4.12& \citenum{kam03a}& 4.85& 4.79 & 4.19 &2.66\\
         PTCDI &3.58*& \citenum{bal20a} &4.24&4.28 & 3.69& 2.18\\
         TPyP &- &- &3.97  & 3.52&2.91 &1.62 
    \end{tabular}
    \label{tab:ea}
\end{table}

The agreement between the HSE06 calculated IPs with experiment is reasonable (Table \ref{tab:ip}) - all calculated values are within 0.5 eV of the experimental values with an RMSE of 0.43 eV. This is slightly higher than the RMSE of previously reported calculated IPs for similar molecular crystals (0.2-0.3 eV),\cite{sch11a,nay09a} however, our sample size is very small (5 molecules in this work compared to 11 and 13 in Refs.~\citenum{sch11a} and \citenum{nay09a} respectively). The HSE06 calculated and experimental IPs follow the same trend, albeit there are only 3 experimental data points so comparison is somewhat limited. HSE PAW overestimates the EA for PTCDI and PTCDA by around 0.75 eV (Table \ref{tab:ea}), so in slightly worse agreement than for IP, though comparison is more limited for the EA as there are only two experimental data points. 

Across all levels of theory, PTCDA has the highest IP and rubrene the lowest, and conversely rubrene has the highest EA and PTCDA the lowest. The HSE06 calculated values in Table \ref{tab:ip} and Figure~\ref{fig:ips} predict that OER should be thermodynamically feasible in PTCDI and TPyP. Rubrene does not appear to have a sufficient oxidation potential to drive OER based on its experimental IP of 4.85 eV, and using the HSE06 IP of 4.32 eV leads to the same conclusion. The HSE06 PAW IP of TBAP lies 0.12 eV below the water oxidation potential, however this difference is within the RMSE of 0.43 eV so OER is conceivable. OER should be possible in PTCDA which has the strongest oxidation potential with experimental IP of 6.6 eV and HSE06 IP of 6.68 eV in good agreement. For PTCDI, based on the experimental IP of 6.1 eV OER should be possible, and although the HSE06 IP is 0.03 eV below the water oxidation potential for PTCDI, this is well within the margin of error. HSE06 predicts water oxidation to be be possible in TPyP with and IP of 6.09 eV.

HSE06 PAW calculations for EA in Table \ref{tab:ea} and Figure~\ref{fig:eas} suggest that HER should be feasible for all molecules apart from PTCDA, whose HSE06 IP of 4.85 eV is on the edge of the margin of error ($\mathrm{IP}(\mathrm{PTCDA})-\mathrm{RMSE}=4.42$eV, proton reduction potential $\epsilon^\mathrm{vac.}_{\mathrm{H^+/H_2}}=4.42$V). Although the experimental IP of PTCDA of 4.12 eV is below the proton reduction potential and would suggest HER to be viable, experimental evidence suggests otherwise as the rate of sacrificial hydrogen evolution is effectively zero in pure PTCDA.\cite{yu24a, guo22a, kam03a} 

Considering gas-phase molecular calculations for the IP and EA, we again notice that energies calculated with B3LYP are the most accurate overall for all the molecular calculations. This is again likely due to a cancellation of errors in molecular orbital energies where stabilising polarisation effects in the solid medium are neglected yet there is a systematic underestimation of the orbital energies when using the hybrid GGA B3LYP functional which compensates for this.\cite{sch11a,kos25a,hsi25a} Gas-phase B3LYP proves to be accurate with respect to higher level periodic HSE06 PAW and experimental data for calculating the IP/EA of the molecular crystals studied here, again suggesting that for simple molecular crystals, computationally inexpensive molecular DFT calculations are sufficient to predict whether HER and OER are thermodynamically feasible in these materials, and therefore are suitable for large-scale computational screening.

\section*{Conclusions}
We have addressed computational descriptors covering the fundamental electronic processes in photocatalytic overall water splitting for a series of organic molecular crystals by both periodic and molecular density functional theory calculations. The optical gaps, IP and EA were calculated for molecular crystals of rubrene, three perylene-based molecules (TBAP, PTCDA, PTCDI) and one porphyrin (TPyP) where we find reasonably good agreement between periodic DFT and experiment for all properties. DFT calculated IP/EA and experimental values taken from the literature both predict that rubrene does not have a sufficiently high oxidation potential to drive OER, and PTCDA has an oxidation potential too high to drive HER. DFT calculations suggest that TBAP, PTCDI and TPyP have the correct hole and electron potentials to drive both OER and HER, suggesting that from an energetic perspective, they would be the most promising molecular starting materials of those tested for the design of a single-component OWS photocatalyst. Moreover, we find that gas-phase molecular calculations with hybrid-GGA functionals such as B3LYP offer comparable accuracy to much more computationally demanding periodic DFT calculations when predicting the optical gap and IP/EA of molecular crystals. We have shown, by comparison to experimental data and the results of high level solid state DFT calculations, that gas-phase molecular DFT/TD-DFT calculations using the B3LYP functional offer sufficient accuracy for screening the optoelectronic properties of molecular crystals for photocatalytic water splitting applications and with a significantly reduced computational cost compared to periodic DFT. In future work, the hopping carrier transport model for organic molecular crystals could be considered, investigating to what extent the computational cost of the hopping-like model could be reduced whilst still predicting qualitative trends in charge transport rates.

\section*{Author Contributions}
The paper was conceived by JDG with advice from KEJ. All computational work and analysis of results was carried out by JDG. The majority of the paper was initially written by JDG. KEJ and DGM assisted with revising the paper. HW and AIC advised on background theory of photocatalytic OWS and organic materials to investigate. DGM and JN advised on the theoretical discussion of charge transport.

\section*{Conflicts of interest}
There are no conflicts of interest to declare.

\bibliography{refs} 

\bibliographystyle{tim}

\end{document}